\documentstyle[12pt,epsf]{article}

\advance \topmargin by -\headheight
\advance \topmargin by -\headsep

\evensidemargin \oddsidemargin
\newcommand{\be}{\begin{equation}}
\newcommand{\ee}{\end{equation}}

\newcommand{\bear}{\begin{eqnarray}}
\newcommand{\eear}{\end{eqnarray}}
\def\ie{{\em i.e.}}
\def\RR{{\rm I\kern-1.6pt {\rm R}}}

\def\ZZ{{\rm Z}\kern-3.8pt {\rm Z} \kern2pt}
\def\IB{\relax{\rm I\kern-.18em B}}
\def\ID{\relax{\rm I\kern-.18em D}}
\def\II{\relax{\rm I\kern-.18em I}}
\def\IP{\relax{\rm I\kern-.18em P}}

\def\np{Nucl. Phys.}
\def\pl{Phys. Lett.}
\def\prl{Phys. Rev. Lett.}
\def\pr{Phys. Rev.}

\def\ijmp{Int. J. Mod. Phys.}

\def\atmp{Adv. Theor. Math. Phys. }
\def\jhep{J. High Energy Phys.}
\def\ptp{Prog. Theor. Phys.}

\def\atmp{Adv. Theor. Math. Phys.}

\begin{document}

\begin{titlepage}

\begin{center} \Large \bf Adding open string modes to the gauge/gravity
correspondence

\end{center}

\vskip 0.3truein
\begin{center}
Alfonso V. Ramallo
\footnote{alfonso@fpaxp1.usc.es}

\vspace{0.3in}

Departamento de F\'\i sica de Part\'\i culas, Universidade de
Santiago de Compostela \\and\\
Instituto Galego de F\'\i sica de Altas Enerx\'\i as (IGFAE)\\
E-15782 Santiago de Compostela, Spain
\vspace{0.3in}

\end{center}
\vskip.5
truein

\begin{center}
\bf ABSTRACT
\end{center}
We review some recent results on the extension of the gauge/gravity correspondence
to include matter in the fundamental representation by adding D-branes to the
supergravity backgrounds. Working in the quenched approximation, in which the
D-branes are considered as probes,  we show how to  compute the meson spectrum
for a general case of brane intersections which are dual to supersymmetric gauge
theories with matter supermultiplets in several dimensions.

\vskip5.6truecm
\leftline{US-FT-2/06}
\leftline{hep-th/0605261 \hfill May  2006}
\smallskip
\end{titlepage}
\setcounter{footnote}{0}

\section{Introduction}	
The gauge/gravity correspondence \cite{jm,MAGOO} provides a closed string
description, based on classical supergravity, of the dynamics of gauge theories at
large 't Hooft coupling. In its first formulation, this duality was applied
to gauge theories in which all  fields are in the adjoint representation.
Obviously, a natural extension of the correspondence would be  the inclusion of
matter (quark) fields in the fundamental representation. This is equivalent to
adding open string degrees of freedom to the supergravity side of the
correspondence and can be achieved by adding D-branes to the supergravity
background. If the number of added D-branes is small as compared to the number of
branes that created the background, we can neglect their backreaction on the
geometry and treat the extra branes as probes. The fluctuations of
these probes correspond to degrees of freedom of open strings  connecting the brane
probe and those that generated the background \cite{KR}. On the field theory side
these open strings are identified with  fundamental  matter multiplets of dynamical
quarks whose masses are proportional to the distance between the two types of
branes. 

In Ref. \cite{KKW} it was proposed to use this setup to add flavor to
some supergravity duals. Indeed, let us consider the intersection of two branes
of different dimensionalities.  In the decoupling limit one
sends the string scale $l_s$ to zero keeping the gauge coupling of the lower
dimensional brane fixed. It is straightforward to see that the gauge coupling of the
higher dimensional brane vanishes in this limit and, therefore, 
the corresponding gauge theory decouples and the gauge group of the higher
dimensional brane becomes the flavor symmetry of the effective theory at the
intersection. In Ref. \cite{KMMW} this approach was applied to the case of the
intersection of D3- and D7-branes which share three common spatial directions
(see also Ref. \cite{D3D7}). If
the number of  D3-branes is large and the number of D7-branes is small one can
employ the probe approximation  and identify the fluctuation modes of the D7-brane
probe in the $AdS_5\times S^5$ geometry with the mesons (or, more rigorously, bound
states) of the gauge theory. Remarkably, the mesonic mass spectrum can be computed
analytically in this case.  Different flavor branes and their spectra for several
backgrounds have been considered in the recent literature (see Refs.
\cite{Sonnen}-\cite{GXZ}). 

In this paper we will first review the analysis of the fluctuation modes
of a brane probe following mostly the approach of Ref. \cite{open}. First of all,
we shall determine a general class of BPS intersections by imposing a
no-force condition. Then, we will particularize our formalism to study the
fluctuation of transverse scalars of the probe in  the background of a stack of
Dp-branes.  In general, the differential equations satisfied by the fluctuations,
and the corresponding mass spectra, must be solved by numerical methods. However,
when the background geometry is $AdS_5\times S^5$ we will show that these equations
can be solved analytically and the mass spectra can be given in closed form. This
exactly solvable cases correspond to the intersections of D3-branes with D7-, D5-
and D3-brane probes, whose induced worldvolume is of the form $AdS_{d+2}\times S^d$
for $d=3,2,1$. As argued in Ref. \cite{KR}, the AdS/CFT
correspondence acts twice in these systems and, apart from 
the holographic description of the four
dimensional field theory on the boundary of $AdS_5$, the fluctuations of 
the  probe are conjectured to be dual to the physics confined to the boundary  of 
$AdS_{d+2}\subset AdS_5$. 
Indeed, the D5- and D3-brane probes in  $AdS_5\times S^5$  are dual to 
${\cal N}=4$, $d=4$ super Yang-Mills theories with codimension one or two defects
and fundamental hypermultiplets localized at the defect. Finally, we will also
briefly review the addition of supersymmetric D-brane probes to the 
$AdS_5\times T^{1,1}$ background, which is dual to an
${\cal N}=1$, $d=4$ superconformal field theory.

\section{BPS Intersections}

\begin{figure}[th]
\centerline{\hskip -.8in \epsffile{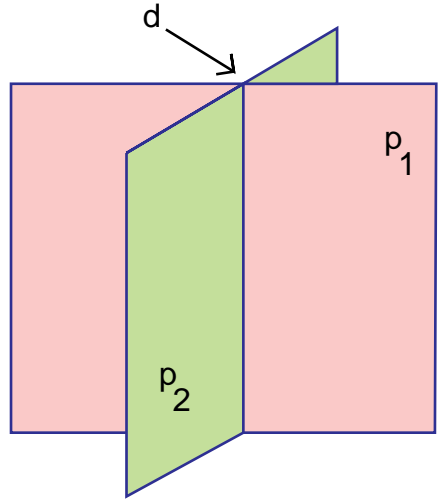}}
\vspace*{8pt}
\caption{An orthogonal intersection of a $p_1$- and a $p_2$-brane along
$d$ spatial directions.\protect
\label{fig1}}
\end{figure}

Let us consider a ten-dimensional background corresponding to a
$p_1$-brane, whose metric can be written  as: 
\begin{equation}
ds^2\,=\,\Biggl[\,{r^2\over R^2}\,\Biggr]^{\gamma_1}\,
dx^2_{1,p_1}
\,+\,
\Biggl[\,{R^2\over r^2}\,\Biggr]^{\gamma_2}\,
d\vec y\cdot d\vec y\,\,,
\label{metric}
\end{equation}
where $dx^2_{1,p_1}$ denotes the $(p_1+1)$-dimensional Minkowski metric, 
$R$, $\gamma_1$ and $\gamma_2$ are constants, 
$\vec y\,=\,(y^1,\cdots, y^{9-p_1})$ and $r^2=\vec y\cdot \vec y$. We
will also assume that the dilaton in such background can be written as:
\begin{equation}
e^{-\phi(r)}\,=\,\Biggl[\,{R^2\over r^2}\,
\Biggr]^{\gamma_3}\,\,,
\label{dilaton}
\end{equation}
where $\gamma_3$ is a new constant. Let us now add to this background a
probe $p_2$-brane sharing $d$ common spatial directions with the
$p_1$-brane. The corresponding orthogonal intersection will be denoted as 
$(d|p_1\perp p_2)$ and is depicted in figure \ref{fig1}. We will assume
that the probe is extended along the directions 
$(x^1,\cdots,x^d, y^1,\cdots, y^{p_2-d})$ and we will denote by $\vec z$
the set of $y$ coordinates transverse to the probe. Notice that $|\vec
z|$ represents the separation of the branes along the directions
transverse to both background and probe branes. Actually, we shall
consider first a static configuration in which the probe is located at a
constant value of $|\vec z|$, namely at $|\vec z|=L$. 
The induced metric on the
probe worldvolume for such a static configuration  will be denoted by:
\begin{equation}
ds^2_{I}\,=\,\,{\cal G}_{ab}\,d\xi^a d\xi^b\,\,,
\end{equation}
with $\xi^a$ being a set of worldvolume coordinates. In what follows we
shall take these coordinates as the common cartesian coordinates
$x^0,\cdots x^d$, together with the spherical coordinates of the 
$y^1,\cdots, y^{p_2-d}$ hyperplane. Actually, assuming that $p_2-d\ge 2$, 
we shall represent the line element of this hyperplane as:
\begin{equation}
(dy^1)^2\,+\cdots (dy^{p_2-d})^2\,=\,d\rho^2\,
+\,\rho^2 d\Omega^2_{p_2-d-1}\,\,,
\label{spherical}
\end{equation}
where $d\Omega^2_{p_2-d-1}$ is the line element of a unit $(p_2-d-1)$-sphere.
It is now straightforward to verify that 
the induced metric $ds^2_{I}$ can be
written as:
\begin{equation}
ds^2_{I}\,=\,\Biggl[\,{\rho^2+L^2\over
R^2}\,\Biggr]^{\gamma_1}\,\, dx^2_{1,d}
\,+\,
\Biggl[\,{R^2\over \rho^2+L^2}\,\Biggr]^{\gamma_2}\,\,(\,
d\rho^2\,+\,\rho^2 d\Omega^2_{p_2-d-1}\,)\,\,.
\label{indmetric}
\end{equation}
The energy density ${\cal H}$ of the probe is determined by its
Dirac-Born-Infeld action. For  static configurations as those we
are considering here, 
${\cal H}\,=\,e^{-\phi}\,\sqrt{-\det {\cal G}}$, where ${\cal G}$ is the
induced metric (\ref{indmetric}). Actually, one can easily prove by using
eqs. (\ref{dilaton}) and (\ref{indmetric})  that
\begin{equation}
{\cal H}\,=\,
\Biggl[\,{\rho^2+L^2\over
R^2}\,\Biggr]^{{\gamma_1\over 2}\,(d+1)\,-\,{\gamma_2\over 2}\,(p_2-d)\,-
\gamma_3}\,\,
\rho^{p_2-d-1}\,\,\sqrt{\det \tilde g}\,\,,
\label{staticH}
\end{equation}
where $\tilde g$ is the metric of the unit $(p_2-d-1)$-sphere. We are
interested in studying BPS intersections, in which the branes do not
exert any force among each other. In our case this no-force condition
requires that ${\cal H}$ be independent of the distance $L$ between the
branes which, in view of the right-hand side of eq. (\ref{staticH}), is
only possible if the number $d$ of common dimensions is related to the
total dimensionality $p_2$ of the probe as:
\begin{equation}
d\,=\,{\gamma_2\over \gamma_1+\gamma_2}\,p_2\,+\,
{2\gamma_3-\gamma_1\over \gamma_1+\gamma_2}\,\,.
\label{BPSrule}
\end{equation}

\section{Dp-brane background}
Let us now particularize our analysis to the case in which the background
geometry is generated by a stack of $N$ parallel Dp-branes. In the string
frame, the near-horizon supergravity solution corresponding to such a
stack has the form displayed in eqs. (\ref{metric}) and (\ref{dilaton}),
with
$p_1=p$, $R$ given by
\begin{equation}
R^{7-p}\,=\,2^{5-p}\,\pi^{{5-p\over 2}}\,\Gamma\Big({7-p\over 2}\Big)\,g_s\,N\,
(\alpha')^{{7-p\over 2}}\,\,,
\label{RDp}
\end{equation}
and with the following  values for the exponents $\gamma_i$:
\begin{equation}
\gamma_1\,=\,\gamma_2\,=\,{7-p\over 4}\,\,,\qquad\qquad
\gamma_3\,=\,{(7-p)(p-3)\over 8}\,\,.\qquad\qquad
\label{Dpgammas}
\end{equation}

Moreover, the Dp-brane solution is endowed with a Ramond-Ramond $(p+1)$-form potential,
whose expression is not relevant in what follows. 
Applying eq. (\ref{BPSrule}) to this background, we get the following
relation  between $d$ and $p_2$:
\begin{equation}
d={p_2+p-4\over 2}\,\,.
\label{Common}
\end{equation}
Let us now consider the case in which the probe brane is another D-brane.
As the brane of the background and the probe should live in the same type II theory, 
$p_2-p$ should be even. Since $d\le p$, we are left with the following
three possibilities $p_2=p, p+2, p+4$, for which eq. (\ref{Common}) gives
$d=p-2, p-1, p$ respectively. Thus, we get the following well-known\cite{rules} set
of orthogonal BPS intersections of D-branes:
\begin{equation}
(p|Dp\perp D(p+4))\,\,,\qquad
(p-1|Dp\perp D(p+2))\,\,,\qquad
(p-2|Dp\perp Dp)\,\,.
\label{BPSlist}
\end{equation}

Let us now study the fluctuations around the $|\vec z|=L$ static
embedding just discussed. Without loss of generality we can take $z^1=L$,
$z^m=0$ ($m>1$) as the unperturbed configuration and consider a
fluctuation of the type:
\begin{equation}
z^1=L+\chi^1\,\,,
\,\,\,\,\,\,\,\,\,\,\,\,\,
z^m=\chi^m\,\,(m>1)\,\,,
\label{perturbation}
\end{equation}
where the $\chi$'s are small. The dynamics of the fluctuations is determined by
the Dirac-Born-Infeld lagrangian which, for the fluctuations of the transverse scalars
we study in this section, reduces to 
${\cal L}\,=\,-e^{-\phi}\sqrt{-\det g}$, where $g$ is the induced metric on the
worldvolume for an embedding of the probe as given by eq.
(\ref{perturbation}). By expanding this lagrangian and keeping up to
second order terms, one can prove that:
\begin{equation}
{\cal L}\,=\,-\,{1\over 2}\,\,\rho^{p_2-d-1}\,\,\sqrt{\det \tilde g}\,\,
\Biggl[\,{R^2\over \rho^2+L^2}\,\Biggr]^{{7-p\over 4}}\,\,
{\cal G}^{ab}\,\partial_{a}\chi^m\,\partial_{b}\chi^m\,\,,
\label{fluct-lag-general}
\end{equation}
where ${\cal G}^{ab}$ is the (inverse of the) metric (\ref{indmetric}).
The equations of motion derived from ${\cal L}$ are:
\begin{equation}
\partial_{a}\,\Bigg[\,{\rho^{p_2-d-1}\sqrt{\det \tilde g}\over 
(\rho^2+L^2)^{{7-p\over 4}}}\,{\cal G}^{ab}\,\partial_{b}\chi
\,\,\Bigg]\,=\,0\,\,,
\label{eom-general}
\end{equation}
where we have dropped the index $m$ in the $\chi$'s. Using the explicit form of
the metric elements ${\cal G}^{ab}$,   eq. (\ref{eom-general})  can be
written as the following differential equation:
\begin{equation}
{R^{7-p}\over (\rho^2+L^2)^{{7-p\over 2}}
}\,\,\partial^{\mu}\partial_{\mu}\,\chi\,+\, {1\over
\rho^{p_2-d-1}}\,\partial_{\rho}\,(\rho^{p_2-d-1}\partial_{\rho}\chi)\,+\,
{1\over
\rho^2}\,\nabla^i\nabla_i\,\chi=0\,\,,
\label{eom-separated}
\end{equation}
where the index $\mu$ corresponds to the directions $x^{\mu}=(t, x^1,\cdots,
x^d)$ and $\nabla_i$ is the covariant derivative on the $(p_2-d-1)$-sphere.
To solve this equation, let us separate variables as:
\begin{equation}
\chi\,=\,\xi(\rho)\,e^{ikx}\,Y^l(S^{p_2-d-1})\,\,,
\label{sepvar}
\end{equation}
where the product $kx$ is performed with the flat Minkowski metric and
$Y^l(S^{p_2-d-1})$ are scalar spherical harmonics on the
$(p_2-d-1)$-dimensional sphere, which satisfy:
\begin{equation}
\nabla^i\nabla_i\,Y^l(S^{p_2-d-1})\,=\,-l(l+p_2-d-2)\,\,Y^l(S^{p_2-d-1})\,\,.
\label{casimir}
\end{equation}
If we redefine the variables as:
\begin{equation}
\varrho\,=\,{\rho\over L}\,\,,
\,\,\,\,\,\,\,\,\,\,\,\,\,\,
\bar M^2\,=\,-R^{7-p}\,L^{p-5}\,k^2\,\,,
\label{newvariables}
\end{equation}
the differential equation (\ref{eom-separated})  becomes:
\begin{equation}
\partial_{\varrho}\,(\varrho^{p_2-d-1}\partial_{\varrho}\,\xi)\,+\,\Big[\,
\bar M^2\,{\varrho^{p_2-d-1}\over (1+\varrho^2)^{{7-p\over 2}}}\,-\,
l(l+p_2-d-2)\varrho^{p_2-d-3}\,\Big]\,\xi\,=\,0\,\,.
\label{fluc}
\end{equation}

\section{$AdS_5\times S^5$ background}
When the background is generated by a stack of D3-branes, the background
geometry is $AdS_5\times S^5$ and the list (\ref{BPSlist}) of BPS
intersections reduces to:
\begin{equation}
(3|D3\perp D7)\,\,,\qquad
(2|D3\perp D5)\,\,,\qquad
(1|D3\perp D3)\,\,.
\label{AdSdefects}
\end{equation}
Remarkably, it turns out that in this case the differential equation
(\ref{fluc}) for the fluctuations of the transverse scalars can be solved
analytically in terms of hypergeometric functions and the spectrum of
values of $\bar M$ can be found exactly. To prove this fact, let us
introduce the quantity $\lambda$, related to the rescaled mass $\bar M$
as:
\begin{equation}
\bar M^2=4\lambda (\lambda+1)\,\,.
\label{Mlambda}
\end{equation}
Then, the solution of (\ref{fluc}) for $p=3$ that is regular when
$\varrho\to 0$ is:
\begin{equation}
\xi(\varrho)\,=\,\varrho^l\,(\varrho^2+1)^{-\lambda}\,
F(-\lambda,  -\lambda+l-1+{p_2-d\over 2}; l+{p_2-d\over 2};-\varrho^2\,)\,\,.
\label{scalarhyper}
\end{equation}
We also want that $\xi$ vanishes when $\varrho\to\infty$. A way to ensure
this is by imposing that
\begin{equation}
-\lambda+l-1+{p_2-d\over 2}\,=\,-n\,\,,
\,\,\,\,\,\,\,\,\,\,\,\,\,\,\,\,\,\,\,\,
n=0,1,2,\cdots\,.
\label{quant}
\end{equation}
When the quantization condition (\ref{quant}) is imposed, the series
defining the hypergeometric function in (\ref{scalarhyper}) truncates,
and the highest power of $\varrho$ is $(\varrho^2)^n$. As a consequence
$\xi$ vanishes as $\varrho^{-(l+p_2-d-2)}$ when $\varrho\to\infty$.
Moreover, notice that the quantization condition (\ref{quant}) of the
values of $\lambda$ implies that the allowed values of $\bar M^2$ are:
\begin{equation}
\bar M^2\,=\,4\Bigg(n+l-1+{p_2-d\over 2}\,\Bigg)
\Bigg(n+l+{p_2-d\over 2}\,\Bigg)\,\,.
\label{exactM}
\end{equation}
Notice that, for the three cases in (\ref{AdSdefects}), $p_2=2d+1$ for
$d=3,2,1$. By using this relation between
$p_2$ and $d$, one can rewrite the mass spectra (\ref{exactM}) of
scalar fluctuations for the  intersections (\ref{AdSdefects}) as:
\begin{equation}
M\,=\,{2L\over R^2}\,\,\sqrt{\Bigg(\,n\,+\,l\,+\,{d-1\over 2}\,\Bigg)
\Bigg(\,n\,+\,l\,+\,{d+1\over 2}\,\Bigg)}\,\,,
\label{generalMS}
\end{equation}
where $M^2=-k^2$ and we have taken into account that, in this case, 
$\bar M^2=-R^4L^{-2}k^2$ (see eq. (\ref{newvariables})). Moreover, in
this $AdS_5\times S^5$ background, the induced metric on
the probe (\ref{indmetric}) can be written as:
\begin{equation}
ds^2_{I}\,=\,{\rho^2+L^2\over R^2}\,\,dx_{1,d}^2\,+\,
{R^2\over \rho^2+L^2}\,d\rho^2\,+\,R^2\,{\rho^2\over \rho^2+L^2}\,\,d\Omega^2_d\,\,.
\label{AdSindmetric}
\end{equation}
It is clear by inspecting (\ref{AdSindmetric}) that, in the conformal
case $L=0$, the metric $ds^2_I$ reduces to that of a product space of the
form $AdS_{d+2}\times S^d$. Interestingly, this same form of the metric is
achieved for $L\not=0$ in the ultraviolet limit $\rho\to\infty$. This
$\varrho\to\infty$ limit is simply the high energy regime of the theory, where the
mass of the quarks, which are proportional  to the brane separation $L$, 
can be ignored and the theory becomes conformal. Therefore, the
$\varrho\to\infty$ behaviour of the fluctuations should provide us  
information about the  conformal dimension $\Delta$ of the corresponding
dual operators. Indeed, in the
context of the AdS/CFT correspondence in $d+1$ dimensions, it is well known that, if the
fields are canonically normalized, the normalizable modes behave at infinity as
$\rho^{-\Delta}$, whereas the non-normalizable ones should behave as
$\rho^{\Delta-d-1}$. In the case in which the modes are not canonically normalized the
behaviours of both types of modes are of the form 
$\rho^{-\Delta+\gamma}$ and $\rho^{\Delta-d-1+\gamma}$ for some $\gamma$.
In our case the fluctuations are given in terms of a  hypergeometric
function in eq. (\ref{scalarhyper}). Taking into account that
for large $\varrho$ the hypergeometric function behaves as:
\begin{equation}
F(a_1, a_2;b;-\varrho^2)\approx c_1\, \varrho^{-2a_1}\,+\,
c_2\, \varrho^{-2a_2}\,\,,\qquad\qquad (\varrho\to \infty)\,\,,
\label{asymptotic-hyper}
\end{equation}
one immediately gets that the associated conformal dimension is:
\begin{equation}
\Delta\,=\,{d+1\over 2}\,+\,a_2-a_1\,\,.
\label{Delta-hyper}
\end{equation}
From the solution (\ref{scalarhyper}) one gets
that $a_1=-\lambda$ and $a_2=-\lambda+l+{d-1\over 2}$. By applying eq.
(\ref{Delta-hyper}) to this case, we get the following value for the dimension of the
operator associated to the scalar fluctuations:
\begin{equation}
\Delta\,=\,l+d\,\,.
\label{generalDeltaS}
\end{equation}
The field theory dual of the three intersections (\ref{AdSdefects}) is
well-known. Indeed, the $(3|D3\perp D7)$ intersection is the case
extensively studied in Ref.~\cite{KMMW} and corresponds in
the UV to an  $AdS_5\times S^3\subset AdS_5\times S^5$ embedding. In this
case the D7-brane is a flavor brane for the ${\cal N}=4$ gauge theory
coupled to an ${\cal N}=2$ fundamental hypermultiplet and the fluctuation
modes of the probe can be identified with the mesons of the theory.  It
turns out that the mass spectra of all the Born-Infeld modes (and
not only those reviewed here that correspond to the transverse scalars)
can be computed analytically\cite{KMMW}. This full set of fluctuation
modes can be accommodated in multiplets, and the mass spectra display the
expected degeneracy. The dual operators in the gauge theory side can be
matched with the fluctuations by looking at the UV dimensions and at the
R-charge quantum numbers. Generically, the dual fields are bilinear in
the fundamental fields and contain the powers of the adjoint fields
needed to construct the appropriate representation of the R-charge
symmetry. 

In the conformal limit $L=0$
the $(2|D3\perp D5)$ intersection represents an $AdS_4\times S^2$ defect
in $AdS_5\times S^5$. The  codimension one dual defect CFT has 
has been studied in detail in Ref.~\cite{WFO}. It corresponds to  
${\cal N}=4$, $d=4$ super Yang-Mills theory coupled to  ${\cal N}=4$,
$d=3$ fundamental hypermultiplets localized at the defect. In Ref.~
\cite{WFO}  the action of the model was constructed and a precise
dictionary between operators of  field theory and fluctuation modes of
the probe was obtained (see also Refs. \cite{EGK,ST}). The meson
spectrum for this defect field theory when the separation of the branes
is non-vanishing has been obtained, for the full set of fluctuation
modes, in Ref. \cite{open}. The corresponding mesonic mass levels can
be obtained analytically and are compatible with the
supermultiplet assignments made in Ref.~\cite{WFO}.

The $(1|D3\perp D3)$ intersection corresponds, in the conformal limit, to
an $AdS_3\times S^1$ defect in $AdS_5\times S^5$ which is  codimension
two in the gauge theory directions\cite{CEGK}.  This system can be
regarded as the dual of  two ${\cal N}=4$ four-dimensional theories
coupled to each other through a bifundamental hypermultiplet living on
the two-dimensional defect. If  a non-zero mass is given to the
hypermultiplet, a mass gap is introduced  in the theory and the mass
spectrum of all modes can   be  obtained in closed form\cite{open}.

\section{Approximate methods for the Dp-brane
background}
When $p\not=3$ the differential equation (\ref{fluc}) cannot be solved
analytically and, thus, we have to make use of other techniques in order
to obtain the fluctuation spectrum. One of these techniques is converting
eq. (\ref{fluc}) into a Schr\"odinger equation of the type:
\begin{equation}
\partial^2_y\,\psi\,-\,V(y)\,\psi\,=\,0\,\,,
\label{Sch}
\end{equation}
where $V$ is some potential. The change of variables needed to transform eq.
(\ref{fluc}) into (\ref{Sch}) is:
\begin{equation}
e^y\,=\,\varrho\,\,,
\,\,\,\,\,\,\,\,\,\,\,\,\,
\psi\,=\,\varrho^{{p_2-d-2\over 2}}\,\,\xi\,\,.
\label{Sch-variables}
\end{equation}
Notice that in this change of variables $\varrho\to\infty$ corresponds to $y\to\infty$,
while the point $\varrho=0$ is mapped into $y=-\infty$. Moreover, 
the resulting potential $V(y)$ takes the form:
\begin{equation}
V(y)\,=\,\Biggl(l-1+{p_2-d\over 2}\Biggr)^2\,-\,\bar M^2\,\,{e^{2y}\over
(e^{2y}\,+\,1)^{{7-p\over 2}}}\,\,.
\label{potential}
\end{equation}
In these new variables, the problem of finding the mass spectrum can be rephrased
as that of finding the values of $\bar M$ such that a zero-energy level for the
potential  (\ref{potential}) exists. By inspecting the form of $V(y)$ one readily
concludes that, for $p<5$, it has a unique minimum at $e^{2y}={2\over 5-p}$. 
Notice that the classically allowed region in the
Schr\"odinger equation (\ref{Sch}) corresponds to the values of $y$ such that $V(y)\le
0$. We would have a discrete spectrum if this region is of finite size or, equivalently,
if the points $y=\pm \infty$ are not allowed classically. When $p<5$ the second
term in (\ref{potential}) vanishes at $y=\pm\infty$ and, thus, only the first term
of V (which is always non-negative) remains in this limit. This means that, indeed,
there is a discrete spectrum of values of $\bar M$. A very useful (and in some
cases accurate) tool to estimate this spectrum is the semiclassical WKB method,
whose starting point is the WKB quantization rule:
\begin{equation}
(n+{1\over 2})\pi\,=\,\int_{y_1}^{y_2}\,dy\,\sqrt{-V(y)}\,\,,
\,\,\,\,\,\,\,\,\,\,\,\,\,\,
n\ge 0\,\,,
\label{WKBquantization}
\end{equation}
where $n\in\ZZ$ and $y_1$, $y_2$ are the turning points of the potential
($V(y_1)=V(y_2)=0$). The WKB method has been very successful
\cite{MInahan,RS} in  the calculation of the glueball spectrum in the context of the
gauge/gravity correspondence. In this method one evaluates the right-hand side of
eq. (\ref{WKBquantization}) by expanding it as a power series in $1/\bar M$ and 
keeping the leading and subleading terms of this expansion. One  obtains  in this
way the expression of $\bar M$ as a function of the principal quantum number $n$
which is, in principle, reliable for large $n$, although in some cases it happens
to give the exact result. The outcome\cite{open} of this analysis is the following 
expression for the  mass levels:
\begin{equation}
M^{WKB}(n,l)\,=\,2\sqrt{\pi}\,{L^{{5-p\over 2}}\over R^{{7-p\over 2}}}\,\,
{\Gamma\Big({7-p\over 4}\Big)\over \Gamma\Big({5-p\over 4}\Big)}\,\,
\sqrt{(n+1)\,\bigg(n\,+\,{7-p\over 5-p}\,\Big(\,l-1\,+\,{p_2-d\over 2}\,\,
\Big)\,\bigg)}\,\,.
\label{WKBmass}
\end{equation}

\begin{table}[!h]
\begin{tabular}[b]{|c|c|c|}   
\hline  
\multicolumn{3}{|c|}{$(2|D2\perp D6)$ with $l=0$}\\
\hline
 $n$  & WKB  & Numerical \\ 
\hline   
\ \ 0 & $11.46$  & $11.34$  \\   
\ \ 1 & $36.67$  & $36.54$  \\   
\ \ 2 & $75.63$  & $75.50$  \\   
\ \ 3 & $128.34$  & $128.20$  \\     
\ \ 4 & $194.80$  & $194.66$  \\
\ \ 5 & $275.01$  & $274.88$  \\ 
\hline
\end{tabular}
\qquad
\begin{tabular}[b]{|c|c|c|}   
\hline  
\multicolumn{3}{|c|}{$(4|D4\perp D8)$ with $l=0$}\\
\hline
 $n$  & WKB  & Numerical \\ 
\hline   
\ \ 0 & $4.31$  & $4.68$  \\   
\ \ 1 & $11.48$  & $11.88$  \\   
\ \ 2 & $21.53$  & $21.94$  \\   
\ \ 3 & $34.45$  & $34.86$  \\     
\ \ 4 & $50.24$  & $50.66$  \\
\ \ 5 & $68.91$  & $69.34$  \\ 
\hline
\end{tabular}
\qquad
\begin{tabular}[b]{|c|c|c|}   
\hline  
\multicolumn{3}{|c|}{$(3|D4\perp D6)$ with $l=0$}\\
\hline
 $n$  & WKB  & Numerical \\ 
\hline   
\ \ 0 & $2.15$  & $1.68$  \\   
\ \ 1 & $7.18$  & $6.78$  \\   
\ \ 2 & $15.07$  & $14.72$  \\   
\ \ 3 & $25.84$  & $25.58$  \\     
\ \ 4 & $39.48$  & $39.34$  \\
\ \ 5 & $55.99$  & $56.02$  \\ 
\hline
\end{tabular}
\caption{Numerical and WKB mass levels for some $(2|Dp\perp D(p+4))$ 
and $(2|Dp\perp D(p+2))$
intersections.}
\label{MassTable}
\end{table}
The mass levels can also be computed numerically by means of the shooting
technique. To apply this technique one first notices that those  solutions  of
eq. (\ref{fluc}) which are regular in the IR behave as $\xi\sim \varrho^{l}$ near
$\varrho\approx 0$, while for large $\varrho$ they  should decrease
as  $\xi\sim \varrho^{-(l+p_2-d-2)}$. In the shooting technique one solves 
numerically the differential equation (\ref{fluc}) for the fluctuations by
imposing the regular behaviour $\xi\sim \varrho^{l}$
at $\varrho\approx 0$ and then one
scans the values of $\bar M$ until the UV behaviour $\xi\sim \varrho^{-(l+p_2-d-2)}$
is fine tuned. This occurs only for a discrete set of values of $\bar M$, which
determines the mass spectrum we are looking for. The numerical values obtained for
some intersections are given in table \ref{MassTable}, where they are compared with
the masses obtained with the WKB mass formula (\ref{WKBmass}).  Notice that,
indeed, for large $n$ the WKB estimate (\ref{WKBmass}) is rather accurate. 

Several comments concerning the meson spectra just found are in order. First of
all,  the masses  for the full set of fluctuations of the intersections 
(\ref{BPSlist}) have been obtained, both numerically and semiclassically, in Ref. 
\cite{open}. Moreover, we
notice from (\ref{WKBmass}) that, for large $n$,  the mass grows linearly with the
excitation number
$n$ (\ie\  $M\sim n$ for $n\to\infty$). This is in contrast with the 
$M\sim \sqrt{n}$ behaviour expected\cite{Schreiber} in confining gauge
theories. Notice also that the mass gap of the theory is just the coefficient
relating $\bar M^2$ and $M^2=-k^2$ in eq. (\ref{newvariables}), namely
$R^{p-7}\,L^{5-p}$ (see also eq. (\ref{WKBmass})). Let us express this quantity in
terms of the quark mass $m_q$ and the Yang-Mills coupling constant $g_{YM}$, given
by:
\begin{equation}
m_q\,=\,{L\over 2\pi\alpha'}\,\,,
\qquad
g^{2}_{YM}\,=\,(2\pi)^{p-2}\,g_s\,(\alpha')^{{p-3\over 2}}\,\,.
\end{equation}
After using the expression (\ref{RDp}) of $R$, one easily gets:
\begin{equation}
R^{p-7}\,L^{5-p}\,=\,
{2^{p-2}\,\,\pi^{{p+1\over 2}}\over \Gamma({7-p\over 2})}\,\,\,
{(m_q)^{5-p}\over g^{2}_{YM} N}\,\,.
\label{massgap}
\end{equation}
It is interesting\cite{MT} to rewrite this
result in terms of the effective dimensionless coupling constant $g_{eff}(U)$ at
the energy scale $U$, which is given by \cite{IMSY}:
\begin{equation}
g_{eff}^2(U)\,=\,g^{2}_{YM}\, N\,U^{p-3}\,\,.
\label{effcoupling}
\end{equation}
It follows from (\ref{massgap}) and (\ref{effcoupling}) that, up to a numerical
coefficient, the mass gap of the theory is:
\begin{equation}
M\,\sim {m_q\over g_{eff}(m_q)}\,\,,
\end{equation}
a result that seems to be universal and coincident with the UV/IR relation found
in Ref. \cite{PP}.

\section{Reduced supersymmetry}
In order to extend the results reviewed in the previous sections to more realistic
scenarios one should consider the addition of branes to less supersymmetric
backgrounds. Here we will mostly concentrate on reviewing the case in which the
five-sphere of the $AdS_5\times S^5$ geometry is substituted by an Einstein space
$X_5$ with less isometries. If, in particular,  $X_5$ is taken as the 
$T^{1,1}$ space we have the so-called Klebanov-Witten (KW) model\cite{KW}, which is
the background corresponding to having a stack of $N$ D3-branes at the tip of the
conifold. The corresponding dual field theory is a  four-dimensional 
${\cal N}=1$ superconformal field theory with gauge group $SU(N)\times SU(N)$ 
coupled to four chiral superfields $A_i$, $B_i$ $(i=1,2)$ in the bifundamental
representation. The ten-dimensional metric for the KW model has the form:
\begin{equation}
ds^2\,=\,{r^2\over L^2}\,dx^2_{1,3}\,+\,{L^2\over r^2}\,dr^2\,+\,
L^2\,ds^2_{T^{1,1}}\,\,,
\label{adspoincare}
\end{equation}
where $L^4={27\over 4}\,\pi g_s N\alpha'{}^2$ and 
$ds^2_{T^{1,1}}$ is the metric of the  $T^{1,1}$ space. This metric can be
written\cite{Candelas} by using the fact that this space can be realized as the
coset
$(SU(2)\times SU(2))/U(1)$ and that 
it is a $U(1)$ bundle over $S^2\times S^2$. Actually, if $(\theta_1,\phi_1)$ and 
$(\theta_2,\phi_2)$ are the standard coordinates of the $S^2$'s and if
$\psi\in [0,4\pi)$ parametrizes the $U(1)$ fiber, the metric 
$ds^2_{T^{1,1}}$ is:
\begin{equation}
ds^2_{T^{1,1}}\,=\,{1\over 6}\,\sum_{i=1}^{2}\,
\big(\,d\theta_i^2\,+\,\sin^2\theta_i\,d\phi_i^2\,)\,+\,
{1\over 9}\,\big(\,d\psi\,+\,\sum_{i=1}^{2}\cos\theta_id\phi_i\,\big)^2\,\,.
\label{t11metric}
\end{equation}
To determine the supersymmetric embeddings of the different D-brane
probes in this background we do not have at our disposal the no-force
argument employed above for the geometries created by D-branes in flat
space. Instead we have to use kappa symmetry, which is based on the fact
that there exists a matrix $\Gamma_{\kappa}$ such that, if $\epsilon$  is
a Killing spinor of the background, only those embeddings obeying
$\Gamma_{\kappa}\epsilon=\epsilon$ are supersymmetric. The matrix 
$\Gamma_{\kappa}$ depends on the embedding
of the probe in the background 
and, therefore, if the Killing spinors of the latter are known, 
the kappa symmetry condition allows to determine such supersymmetric
embeddings in a systematic way. This analysis has been carried out in
detail in Ref. \cite{acr} and it will be summarized here.
First of all, let us recall that the $T^{1,1}$ space is topologically 
$S^2\times S^3$. Thus, there is the possibility of having D3-branes
wrapping a three-cycle of $T^{1,1}$, which allows for a rich zoology of BPS
intersections.  Moreover, the $T^{1,1}$ is a
Sasaki-Einstein space and therefore, its  six-dimensional cone 
$C(T^{1,1})$ with metric $dr^2\,+\,r^2\,ds^2_{T^{1,1}}$ is a Calabi-Yau
manifold with complex dimension three. A very convenient set of
complex coordinates of $C(T^{1,1})$ is the following:
\begin{equation}
z_i\,=\,\tan \bigg({\theta_i\over 2}\,\bigg)\,e^{i\phi_i}
\,\,\,(i=1,2)\,\,,\qquad
z_3=r^3\sin\theta_1\sin\theta_2\,e^{-i\psi}\,\,. 
\label{cc}
\end{equation}
It turns out that there exists a family of three-cycles 
${\cal C}_3\in T^{1,1}$ such that  a D3-brane located at the center of
the $AdS_5$ space and wrapping ${\cal C}_3$ is $1/8$ supersymmetric.
In terms of the complex coordinates (\ref{cc}) these cycles can be
simply characterized  as the interesction of 
the locus of the polynomial equation
\begin{equation}
z_1^{m_1}\,z_2^{m_2}\,=\,{\rm constant}\,\,
\label{threecycles}
\end{equation}
in the cone $C(T^{1,1})$ with its $T^{1,1}$ base.  
To identify the field theory dual of the D3-branes wrapping the
three-cycles (\ref{threecycles}) one must take into account that, according to the
standard AdS/CFT arguments,  the volume of the cycle determines the
conformal dimension of the dual field theory operator. Moreover, in the
KW theory there is a $U(1)$ baryon number symmetry under which the 
$A_i$ ($B_i$) fields have baryon number $+1$($-1$). On the
gravity side of the AdS/CFT correspondence, the baryon number 
can be identified with the third homology class of the three-cycle 
${\cal C}_3$ over which the D3-brane is wrapped. By using these facts one
can verify that these D3-branes are dual to dibaryonic operators built out
from the $A_i$ and $B_i$  fields. For example\cite{GK}, the D3-brane wrapping the
three-cycle with $m_1=1$, $m_2=0$ is dual to an operator of the form
$\det (A)$, whereas for arbitrary values of $m_1$ and $m_2$ it
corresponds to an operator with higher baryon number. It is also
interesting to point out that one can study\cite{BHK} the fluctuations of the
D3-branes, which correspond to mesonic excitations of the dibaryon
operators. 

One can also find supersymmetric embeddings of D5-branes which give rise
to defect theories, similar to the $(2|D3\perp D5)$ intersections in flat
space studied above. In this case the D5-brane must be extended along
two Minkowski coordinates, as well as in the holographic coordinate $r$,
and wrap a two-cycle  ${\cal C}_2$ of $T^{1,1}$, which can be simply
characterized by the equations:
\begin{equation}
\theta_2\,\,=\,\theta_1\,\,,\qquad
\phi_2\,=\,2\pi-\phi_1\,\,,
\end{equation}
with the coordinate $\psi$ being constant. 

The flavor branes of the KW model are D7-branes  filling the four
spacetime Minkowski directions $x^{\mu}$ and extended along a certain
submanifold ${\cal M}_4$ of the conifold $C(T^{1,1})$. In terms of 
the complex coordinates (\ref{cc}) this submanifold ${\cal M}_4$ can be 
described by the following polynomial equation 
\begin{equation}
z_1^{m_1}\,z_2^{m_2}\,z_3^{m_3}\,=\,{\rm constant}\,\,.
\label{D7-T11}
\end{equation}
The meson spectrum for the flavor D7-branes embedded on the conifold geometry as in
eq. (\ref{D7-T11}) has been computed in Ref. \cite{Ouyang} (see also Ref.
\cite{Kuper}).

The same methodology reviewed here for the conifold has been
applied in Refs. \cite{CEPRV} and \cite{CER} to the metrics of the form 
$AdS_5\times Y^{p,q}$ and $AdS_5\times L^{a,b,c}$, where $Y^{p,q}$ and
$L^{a,b,c}$ are the recently discovered\cite{GMSW2} five dimensional Sasaki-Einstein
spaces. Moreover, the kappa symmetry condition $\Gamma_{\kappa}\epsilon=\epsilon$
has also been employed in Refs. \cite{flavoring} and \cite{CPR} to determine
the supersymmetric embeddings of D-branes in the confining
Maldacena-N\'u\~nez\cite{MN} background (for a review on the extension of the
gauge/gravity duality to confining theories, see Ref. \cite{EP}).

\section{Concluding Remarks}
In this paper we have reviewed the holographic description of  mesons by using the
fluctuation of brane probes in supergravity backgrounds. We have restricted
ourselves to  describe the simplest configurations with a large amount of
supersymmetry. For models more suitable for the phenomenological description of QCD
the interested reader should consult, specially, Refs. 
\cite{Johana}, \cite{KMMW-two} and \cite{Sakai}. Let us finish by
mentioning that there have been  different attempts to go
beyond the probe approximation and to obtain a backreacted supergravity solution
\cite{Beyond}.

\section*{Acknowledgments}

This work is based on talks given by the author at the ``Workshop on gravitational
aspects of strings and branes" held at Miraflores de la Sierra (Madrid, Spain) in
January 2005 and at Oxford University in October 2005. I am grateful to the
organizers of the workshop and to Martin Schvellinger for the opportunity to present
this material.  Discussions with D. Arean, F. Canoura, J. D. Edelstein, C.
N\'u\~nez, D. Rodriguez Gomez and M. Schvellinger are gratefully acknowledged.
This work  was
supported in part by MCyT, FEDER and Xunta de Galicia under grant
FPA2005-00188 and by  the EC Commission under  
grants HPRN-CT-2002-00325 and MRTN-CT-2004-005104.

\end{document}